\documentclass[12pt]{article}
\usepackage[english]{babel}
\usepackage{amsmath}
\usepackage{times}
\usepackage[body={6.5in,9.5in}]{geometry}

\usepackage{amsmath}
\usepackage{graphicx}
\usepackage{subfigure}
\usepackage{color}

\title{\it  Lorentz invariant ``potential magnetic  field'' \\and  magnetic flux  conservation
\\  in  an ideal 
relativistic plasma}
\author{{F. Pegoraro\footnote{\tiny e-mail $~~~$ francesco.pegoraro@unipi.it}}\\
{\small \it  Physics Department, Pisa University} \\{\small  \it  largo Pontecorvo 3, 56127 Pisa, Italy}}
\date{}

\begin{document}
\maketitle

\begin{abstract}

Lorentz invariant   scalar functions of the magnetic field  are defined  in an ideal relativistic plasma.  These invariants are advected by the plasma fluid motion  and  play the role of the {\it potential magnetic field} introduced   by R. Hide in  {\it Ann. Geophys.},  {\bf 1}, 59 (1983) on the line of Ertel's theorem. From these invariants we recover  the Cauchy conditions for the magnetic field  components in the Eulerian-Lagrangian variable mapping.  In addition the adopted procedure allows us to  formulate Alfv\`en theorem for the conservation of the magnetic flux through a surface 
comoving  with the plasma in a Lorentz invariant form.

\end{abstract}

\section{Introduction}

The nonlinear dynamics of relativistic plasmas is presently under extensive theoretical and experimental investigation, both in the context of laboratory plasmas such as laser-produced
plasmas and in the context of high-energy astrophysics.
While  the description of the relativistic plasma dynamics  would  require  a fully kinetic treatment 
involving the relativistic Vlasov equation coupled to  Maxwell's equation,
 on  the large spatial and temporal scales  that are of interest for astrophysical plasmas fluid type approximations can be usefully  adopted. This is in particular the case in physically  complex settings such as high energy plasmas in curved space time, see e.g. in  the recent article \cite{Zan} for the case of magnetized neutron stars and  pulsar winds. Relativistic fluid descriptions have also been  used  in a simplified modelling   of magnetic reconnection in high energy plasmas, see e.g., \cite{Zen}. 

A common  feature of these descriptions,  in the  limit where   dissipation and microscopic effects are disregarded,  is the occurrence of topological invariants, such as the conservation of the magnetic flux through a surface comoving with the plasma (Alfv\`en theorem) in ideal Magnetohydrodynamics, that  restrict the plasma dynamics. In fact the process of magnetic reconnection mentioned above arises from the local violation 
of these constraints due to the local violation of the ideal  Ohm's law, that is of the  condition ${\bf E} +({\bf v} /c) \times  {\bf B} = 0$, where ${\bf E}$ and ${\bf B} $ are the electric and magnetic field and ${\bf v}$ is the fluid plasma velocity.

 In general, even for relativistic plasmas,  these  topological constraints are formulated in a  form that is not explicitly invariant under Lorentz transformations even if the  ideal plasma condition ${\bf E} +({\bf v} /c) \times  {\bf B} = 0$ is fully relativistic and covariant, as can be seen explicitly by rewriting it  in the 4 dimensional form 
 ${\bf  F}_{\mu\nu} {\bf u}^\nu =0$ where ${\bf  F}_{\mu\nu}$ is the electromagnetic field tensor and ${\bf u}^\nu$ is the plasma fluid 4-velocity. \\  In order to bypass  this limitation 
in two previous papers  \cite{Peg1,Peg2}  the concept of  covariant magnetic connections between fluid elements has been introduced.  A basic ingredient in the definition of invariant connections is a procedure of time resetting along the fluid  trajectories that is  compatible with the ideal Ohm's law and that is required in order to restore the simultaneity between  two fluid elements at different  spatial locations that is  not preserved  by a Lorentz transformation.

Following a different  angle of approach, but under  the same ideal plasma  assumption,  R. Hide defined \cite{hide1,hide2}, by analogy to the potential vorticity introduced by Ertel \cite{ertel} for  fluids,  a  {\it potential magnetic field}   that is advected by the plasma velocity field.

In the present article we provide  a relativistic definition  of the potential magnetic field that is explicitly Lorentz invariant.  This generalization turns out to be rather  convenient  as  it  allows us on the one hand to  recover  the well known Cauchy conditions\footnote{This name is a generalization to a magnetic field of a term  originally used for the  vorticity in an incompressible fluid \cite{frish}.} for the magnetic field  components in the Eulerian-Lagrangian variable mapping  (see \cite{newcomb})  and on the other to prove a Lorentz invariant form of the Alfv\'en theorem that, as in the case of the covariant connections in \cite{Peg1,Peg2},  requires a time resetting procedure in order to restore simultaneity among the different points of the comoving surface theough which the magnetic flux is computed.

This article is organized as follows. In Sec.\ref{General}, following closely   \cite{Peg2},  we recall  for the sake of self-containedness the basic features of the Lichnerowicz-Anile (LA) representation  \cite{Li,Ani} of the electromagnetic field tensor ${\bf  F}_{\mu\nu} $, the definition of the magnetic   4-vector ${\bf b}^\mu$ and of the electric   4-vector ${\bf e_\mu}$, the ideal plasma limit where the electric 4-vector ${\bf e_\mu}$ vanishes  and the expression  of the divergentless dual tensor  ${\bf G}^{\mu\nu}$ in this ideal limit in tems of the 4-vectors ${\bf b^\mu}$ and ${\bf e^\mu}$.   Then  we recall  the gauge freedom  (see  also \cite{D'A}) in the definition of the magnetic 4-vector. This gauge freedom plays a very important role in the derivation of the potential magnetic field and of the Alfv\`en theorem as it allows us to move easily  from a Lorentz invariant  
4-dimensional formulation to a  3+1-dimensional  formulation in a chosen  frame and back, whichever formulation is more convenient  through  the different steps of the proofs.
In Sec.\ref{intr} we derive the explicit relativistic invariant expression of the potential magnetic field. In Sec.\ref{3+1}   using the gauge freedom mentioned above,  we show that  the relativistic  expression of the potential magnetic field  reduces to   Hide's definition when expressed in a 3+1-dimensional form. In Sec.\ref{IAFT},  by introducing appropriately conserved ``charges''  that are closely  related to the  potential magnetic field defined in Sec.\ref{intr} and by using the gauge freedom explicitly, we recover in a Lorentz invariant form the Alfv\`en theorem for the conservation of the magnetic flux through a surface comoving with the plasma.  Finally in Sec.\ref{concl}  the main results of this article are summarised and possible extensions are indicated.

\section{Electric and magnetic 4-vectors}\label{General}

Following  \cite{Peg2,Peg3} we adopt the   so called Lichnerowicz-Anile (LA) representation  \cite{Li,Ani} (see also \cite{D'A}) of the relativistic  e.m. field tensor  ${\bf F}_{\mu\nu}$
\begin{equation}\label{2} {\bf F}_{\mu\nu}  =  {\varepsilon_{\mu\nu\lambda\sigma} {\bf b}^{\lambda}{\bf u}^{\sigma}}\,  + \,  {[{\bf u}_{\mu}{\bf e}_{\nu} - {\bf u}_{\nu}{\bf e}_{\mu}]} \, , \end{equation}
where  ${\bf b}^\mu$  is the {\it 4-vector magnetic  field} and ${\bf e}_\mu$  is the {\it 4-vector electric field}, 
with  ${\bf  u}^\mu {\bf  e}_\mu =0$ and ${\bf u}_\mu{\bf  b}^\mu =0$. The  4-vectors ${\bf  e}_\mu$ and ${\bf  b}^\mu$ are related to the standard electric and magnetic fields ${\bf E}$ and ${\bf B}$ in 3-D space by
\begin{equation}\label{2-b}
{\bf  b}^\mu = \gamma({\bf B}  + {\bf E} \times {\bf v} \,  , \, {\bf B} \cdot  {\bf v}\  ), 
\end{equation}
and 
\begin{equation}\label{2-c}
{\bf  e}_\mu = \gamma({\bf E} + {\bf v} \times {\bf B} \,  , \, -{\bf E} \cdot  {\bf v}   ) ,
\end{equation}
with ${\bf  e}_\mu {\bf  b}^\mu =  {\bf E}  \cdot {\bf B}$. \, We have adopted  the Minkowski metric tensor $\eta_{\mu\nu}$ defined by $(+,+,+,-)$ and  normalized 3-D velocities ${\bf v}$ to the speed of light:  $\gamma$ is the relativistic Lorentz factor and we have used ${\bf u}^\mu = \gamma ( {\bf v},1)$ and $~{\bf u}_\mu {\bf  u}^\mu = -1$. The orthogonality conditions  ${\bf u}^\mu {\bf e}_\mu = {\bf  u}_\mu {\bf  b}^\mu =0$  make the LA  representation  unique.\\
The LA representation  allows us to separate  covariantly  the magnetic and the  electric  parts of the e.m. field tensor relative to a given  plasma element moving  with  4-velocity ${\bf u}^\mu$.  In the  local rest frame of this plasma element   the time components of ${\bf e}_\mu$ and of ${\bf  b}^\mu$ vanish, while their space components reduce to the standard 3-D electric and magnetic fields. 
\\
A corresponding representation holds for the dual tensor ${\bf G}^{\mu\nu} \equiv  \varepsilon^{\mu\nu\alpha\beta} {\bf F}_{\alpha\beta}/2$ with $ {\bf e}_\mu $ and $ {\bf b}^\mu $ interchanged. \, Thus:
\begin{equation}  {\bf G}^{\mu\nu}  =  {\varepsilon^{\mu\nu\lambda\sigma} {\bf u}_{\lambda}{\bf  e}_{\sigma}}\,  + \,  {[{\bf u}^{\mu}{\bf  b}^{\nu} -{\bf   u}^{\nu}{\bf  b}^{\mu}]},  \quad  {\rm with} \,
{\bf  e}_\mu = {\bf  F}_{\mu\nu} {\bf  u}^\nu  \quad {\rm and} \quad  {\bf  b}^\mu = {\bf G}^{\mu\nu}  {\bf  u}_\nu.  \label{defin}\end{equation}

If the ideal Ohm's law ${\bf  F}_{\mu\nu} {\bf u}^\nu =0$   holds, the electric 4-vector ${\bf  e}_\mu$ vanishes, the tensors ${ \bf F}_{\mu\nu}$ and ${\bf G}^{\mu\nu}$ have rank two   and can be written as 
\begin{equation}\label{2b} {\bf F}_{\mu\nu}  =  {\varepsilon_{\mu\nu\lambda\sigma} {\bf  b}^{\lambda}{\bf  u}^{\sigma}}\, , \qquad  \,  {\bf G}^{\mu\nu}  = {[{\bf  u}^{\mu} {\bf b}^{\nu} - {\bf  u}^{\nu} {\bf b}^{\mu}]} \, , \end{equation}
with 
\begin{equation}\label{2c} { \bf F}_{\mu\nu} {\bf b}^\nu =  {\bf F}_{\mu\nu} {\bf u}^\nu  =0,  \ \end{equation} 
\begin{equation}\label{2d}  {\bf F}_{\mu\nu} {\bf G}^{\nu\mu} = 0 \,  \rightarrow   {\bf E}  \cdot {\bf B} =0, \qquad {\rm and} \quad  {\bf  b}_\mu{\bf  b}^\mu  =  {\bf G}_{\mu\nu}  {\bf G}^{\nu\mu}/2  = {\bf F}_{\mu\nu}  {\bf F}^{\nu\mu}/2  . \end{equation}
In this case we can use ${\bf  e}_\mu =0$ in order to express ${\bf  b}^\mu$ in terms of $ {\bf B}$ and ${\bf v}$ only as
\begin{equation}\label{b}  \  {\bf b}^{\mu} = \gamma ( {{\bf B}}/{\gamma ^2}  + {\bf  v} \, ({\bf v} \cdot {\bf B}) \,  ,\, {\bf v}\cdot{ \bf B}) .\end{equation}
Note that in  general   $\partial_\mu {\bf b} ^\mu \not =0$ while from Maxwell's equations  we have 
\begin{equation}\label{div}  \partial_\mu  {\bf G}^{\mu\nu} = 0. \end{equation}

\subsection{Gauge freedom}

As shown in detail in \cite{Peg2,D'A} a gauge freedom is allowed in the definition of the magnetic 4-vector  field ${\bf b}^\mu$ in the LA representation  provided  we relax the orthogonality condition ${\bf b}^\mu {\bf u}_\mu = 0$:
\begin{equation}\label{gauge}    {\bf b}^\mu \rightarrow  {\bf h}^\mu \equiv {\bf b}^\mu +  g \, {\bf u}^\mu ,  \end{equation}
where  $g$ is a  free scalar field and the velocity 4-vector ${\bf u}^\mu$ satisfies the  continuity equation 
\begin{equation}\label{cont} \partial_\mu(N {\bf u}^\mu) =0,   \end{equation} 
with   $N $ is the proper density of the plasma element and $N {\bf u}^\mu$ of the  density 4-vector.
\,
Different choices of  the gauge field $g$ allow us to impose  specific conditions on ${\bf h}^\mu$.
If we take in a given frame\footnote{Actually this gauge is Lorentz invariant  since the quantity $ -  {\bf v}\cdot{ \bf B} $ can be written as a Lorentz scalar. Its expression in a frame moving with respect to the chosen frame with velocity 4-vector $V_\mu$ is
given by  $-  (V_\mu b^\mu)/(V_\nu u^\nu)$. } the {\it magnetic gauge} 
\begin{equation}\label{gauge2}  g   = -  {\bf v}\cdot{ \bf B} , \end{equation}
we can make the time component of ${\bf h}^\mu$ vanish and   ${\bf h} \, || \, { \bf  B}$ in that frame i.e., \begin{equation} \label{hmu} {\bf h}^{\mu} =  ( {{\bf B}}/{\gamma } \,  ,\, 0) . \end{equation}
\, Note that the expression for ${\bf G }^{\mu\nu}$ in Eq.(\ref{2b}) is unchanged if we insert ${\bf h}^\mu$ for ${\bf b}^\mu$ in Eq.(\ref{2b}).

\section{Advected relativistic ``potential magnetic field''}\label{intr}

Let $S $ be a scalar function in Minkowski space-time, then
\begin{equation}\label{1}  
(\partial_\nu S)  \, \partial_\mu {\bf G}^{\mu\nu} =  \partial_\mu  [ (\partial_\nu S)  \, {\bf G}^{\mu\nu}] -   (\partial_\mu \partial_\nu S) \,  {\bf G}^{\mu\nu} = \partial_\mu  [ (\partial_\nu S)  \, {\bf G}^{\mu\nu}]  = 0. \end{equation}
This is a generalization of the solenoidal property of the magnetic field  (for $S = t$, \, Eq.(\ref{1})  reduces to $\nabla \cdot {\bf B} = 0$) and includes the induction equation whose components are recovered for $S \, = \, x,\, y, \, z,$ respectively.
\\ In the ideal MHD case from Eq.(\ref{2b})  we obtain 
\begin{equation}\label{4}  \partial_\mu  [ (\partial_\nu S)  \, {\bf G}^{\mu\nu}]  =  -   \partial_\mu  [ {\bf  b}^\mu \partial_\tau S - {\bf u}^\mu {\bf b}^\nu \partial_\nu S] =  0.\end{equation}
We find it convenient to choose $S$ such that it is advected by the 4-velocity field $u^\mu$  (e.g. any function of the initial spatial positions ${\bf a}$ at $t=0$). Then 
	 \begin{equation}\label{5}  \partial_\tau S = {\bf u} ^\mu  \partial_\mu  S =  0,\end{equation}
and
\begin{equation}\label{6}  \partial_\mu  [ {\bf u}^\mu {\bf b}^\nu \partial_\nu S] =  0.\end{equation}
Using Eq.(\ref{cont}),  from Eq.(\ref{6})  we obtain
\begin{equation}\label {7}  \partial_\mu  [ (N/N) {\bf u}^\mu {\bf  b}^\nu \partial_\nu S] =N \, \partial_\tau [ ({\bf  b}^\nu \partial_\nu S)/N ] =  0.\end{equation}
Equation (\ref{7}) is gauge invariant under  ${\bf  b}^\mu \to {\bf  b} ^\mu + g \,  {\bf  u}^\mu$ for any scalar function $g$ since $\partial_\tau S =  0.$

The scalar quantity  $ ({\bf  b}^\nu \partial_\nu S)/N$ is a relativistic Lorentz invariant generalization of the {\it potential magnetic field} introduced by R. Hide in \cite{hide1,hide2} along the lines of the potential vorticity and of Ertel theorem \cite{ertel}
in fluid dynamics.

\section{  3+1 formulation in a  chosen reference frame} \label{3+1}

From $\partial_\tau [ ({\bf b}^\nu \partial_\nu S)/N ] =  0$ and  $\partial_\tau  S  =  0$, we obtain
\begin{equation}\label {8}   ({\bf b}^\nu \partial_\nu S)/N =   {\cal K}(S,{\bf  a}),   \end{equation}
where ${\bf  a}$ gives  in the chosen reference  frame the spatial initial conditions (at $t = t_o$) of the trajectories of the plasma fluid element, i.e.,   the 3D Lagrangian coordinates, and ${\cal K}(S, {\bf  a}) $ is  a function of the initial conditions and of the advected scalar $S$
that we are free to chose as convenient.

Using  the gauge freedom (\ref{gauge}) in the form given by  Eq.(\ref{hmu})  we can rewrite   Eq.(\ref{8}) as 
\begin{equation}\label {9}    \frac{ { \bf B}^i }{\gamma N} \, \frac{\partial {\bf a}^j}{ \partial {\bf x}^i }\, \frac{ \partial S} { \partial {\bf a}^j}  =   {\cal K}(S, {\bf a}). \end{equation}

Choosing $S=  {a}^k$  with $k = 1,2,3$, respectively,  and recalling that $\gamma N$  is the density  $n(x,t) = n_o({\bf a})/J({\bf a},t) $  in 
3-D  space where $J({\bf a},t)   = det |\partial{\bf  x}/\partial {\bf a}| $ is the 3-D Jacobian determinant,  with self-explaining notation,  for each choice  $S=  {a}^k$ we obtain 
\begin{equation}\label {10}    \frac{ { \bf B}^i }{\gamma N} \, \frac{\partial {\bf a}^k}{ \partial {\bf x} ^i } =   {\cal K}^{k} ({\bf a}),  \end{equation}
i.e. 
\begin{equation}\label {11}   { \bf B}^i ({\bf x},t) =  \, \frac{\partial {\bf x}^i}{ \partial {\bf a}^k} \,  \frac{ n_o({\bf a}) \, {\cal K}^{k} ({\bf a})}{J({\bf a},t)} =    \, \frac{\partial {\bf x}^i}{ \partial {\bf a} ^k} \,  \frac{{\bf B}_{o}^{k} ({\bf a})}{J({\bf a},t) },  \end{equation}
which corresponds to the ``Cauchy condition''  given by   Eq.(2.22) of  \cite{newcomb} (see also Eq.(19) of \cite{AMP-13}) that  can be derived  from the conservation of the  flux  of the 3D magnetic field ${\bf B}$ through a surface comoving with an ideal plasma.

 Note that, knowing ${\bf u}^\mu({\bf x},t)$ in terms  of ${\bf u}^\mu({\bf a}, t_o)$ and imposing  ${\bf b}^\mu {\bf u}_\mu = 0$ for all $t$,  it is possible to reverse the gauge transformation both on the r.h.s. (using Eq.(\ref{2}) with ${\bf e}_\mu =0$) and on the l.h.s. of Eq.(\ref{11})
(using Eq.(\ref{b})) and obtain the Cauchy condition for the 4-vector ${\bf b}^\mu$ in agreement with Eq.({\ref8}).  
 
\section{Invariant Alfv\`en flux theorem}\label{IAFT}

\subsection{Preliminary proposition}\label{irish}

We will make use of  the following result  (see e.g. \cite{Fitz}): if a 4-vector field ${\bf  Q}^\mu$ satisfies the continuity equation $\partial_\mu {\bf  Q}^\mu  =0$ and if its component vanish outside a finite spatial region, then the ``charge'' ${\cal Q} $ 
defined as the space integral of its time component  ${\bf  Q}^0$ 
\begin{equation}  \label{alf1}  
{\cal Q}  = \int d^3x\, {\bf  Q}^0 , 
\end{equation} 
is constant in time and is a Lorentz invariant. We will make use of this result in combination with Eq.(\ref{6}) for different choices of the advected scalar $S$ in order to rephrase  in a Lorentz invariant framework the  magnetic flux conservation (Alfv\`en theorem) 
that is usually proven in a fixed frame. .

 \subsection{Flux conservation through a comoving 2D surface}\label{FS}

First  we take 
\begin{equation}  \label{alf4}  
{\bf  Q}_{\cal A}^\mu = {\bf u}^\mu {\bf b}^\nu \partial_\nu S_{\cal A} ,
\end{equation} 
where the  function $S_{\cal A}$ is defined in a given frame at $t = 0$ in terms of  the characteristic function  of a finite size,  disk-like  domain bounded 
 by  a 2D spatial surface  ${\cal S}({\bf a})$ of the initial conditions (the base of the disk), by the corresponding   surface displaced by the infinitesimal shift 
$ \Delta  {\bf x}^\mu =  {\bf  u}^\mu \Delta \tau $ (the top of the disk)
and by the ``ribbon'' connecting the  rims of the two surfaces (the side of the disk).
The function  $S_{\cal A}$ is advected by the flow 4-velocity ${\bf u}^\mu$, 
i.e.  for all times $t$  in the chosen frame $S_{\cal A} = 1$ inside the advected domain 
and   $S_{\cal A} = 0 $ outside.  \\ Since  ${\bf  Q}_{\cal A}^\mu$ is left invariant\footnote{Note in passing that the 4-vectors 
${\bf  Q}_{\cal A}^\mu$ and  ${\bf  Q}_{\cal S}^\mu$,
 see below,  lie on the two-dimensional ``connection hypersurfaces'' defined in \cite{Peg2}. This indicates that the latter formalism could have been also adopted in the present analysis.} by the  gauge transformation (\ref{gauge2}) it can be rewritten as 
\begin{equation}  \label{alf5}  
{\bf  Q}_{\cal A}^\mu = ( {\bf u}^\mu/\gamma)\,  {\bf B}^i  \partial_i S_{\cal A} \, , \qquad {\rm so \, \, that } \quad {\cal Q}_{\cal A} = \int d^3x\, {\bf B}^i  \partial_i S_{\cal A} =0 . \end{equation} 
The last equality in Eq.(\ref{alf5}) follows from the fact that ${\bf B}$ (which is taken at $t=0$) is divergence-free and that, because of the choice of the function $S_{\cal A}$, the space integral reduces to the surface integral over the closed surface  $\partial{\bar  {\cal A}}$ delimiting the spatial projection  ${\bar  {\cal A}} $ of the domain ${{\cal A}} $ (only spatial derivatives are present in Eq.(\ref{alf5}) because of the gauge used).
\\ For the sake of clarity  we will now refer to a 3+1 notation  and  use  the fields  ${\bf v}$,  ${\bf B}$ and ${\bf E}$  so as to make  the correspondence with the standard derivation of the  Alfv\`en theorem explicit. There are three contributions to the vanishing flux in Eq.(\ref{alf5}): the flux through the base surface,  that  through  the ribbon and that  through  the top surface.
From the space components  of  $\partial_i S_{\cal A} $ at the ribbon and  using   ${\bf E}= { \bf B} \times  ({\bf v}/c )$  we see that the flux  of ${\bf B} $ through  the ribbon equals  $\Delta t = \gamma \Delta \tau$ times the circulation of ${\bf E}$ along the rim of the domain ${\bar  {\cal A}}$. By virtue of the induction equation this flux cancels the difference between the flux of ${\bf B} (t)$ and  that of ${\bf B} (t + \Delta t) \sim {\bf B} (t )  + [\partial_t {\bf B} (t ) ]\,  \Delta t $  through the top surface. Thus Eq.(\ref {alf5})  implies that the flux  of  ${\bf B} (t =0)$ through the base surface (having inverted the direction of the normal to  this surface) and that of ${\bf B} (t =\Delta t)$ through the same surface shifted by  $\Delta {\bf x} = {\bf v} \Delta t$ (the top surface) are the same. This equality, being valid for all times $t$, recovers the flux conservation theorem in an ideal plasma and makes it Lorentz invariant  by virtue of the proposition in Sec.\ref{irish}.

Finally we note when the domain 
 ${\cal A} $ is Lorentz  boosted  to a moving frame the   points on the boosted base surface (and similarly for the  points on the top surface) are no longer simultaneous   because  simultaneity between points that are spatially  separated is not preserved by a Lorentz transformation. 
 This lack of simultaneity in the boosted frame can be corrected, exploiting the flux conservation result derived above  for each infinitesimal portion of the base and top sufaces, by a procedure of ``time resetting'' along the trajectories of the 4-velocity ${\bf u}^\mu$  that follows the method adopted in   \cite{Peg2,Peg1},  see section below Eq.(9) in \cite{Peg1} or Sec.5.1 in \cite{Peg2}.

\subsection{Invariant vanishing of the flux through an advected closed  2D surface }

Here  take 
\begin{equation}  \label{alf2}  
{\bf  Q}_{\cal D}^\mu = {\bf u}^\mu {\bf b}^\nu \partial_\nu S_{\cal D} ,
\end{equation} 
where the  function $S_{\cal D}$ is defined in terms of  the characteristic function  of a finite spatial domain ${\cal D}({\bf a})$ of the initial conditions. \\
The derivation now follows  with only minor changes the  one in Sec.(\ref{FS}) and we   rewrite ${\bf  Q}_{\cal D}^\mu $ as 
\begin{equation}  \label{alf3}  
{\bf  Q}_{\cal D}^\mu = ( {\bf u}^\mu/\gamma)\,  {\bf B}^i  \partial_i S_{\cal D} \, , \qquad {\rm so \, \, that } \quad {\cal Q}_{\cal D} = \int d^3x\, {\bf B}^i  \partial_i S_{\cal D}  \, = 0. \end{equation} 

We take  the simple case where $\partial {\cal D}$ is isomorphic to a sphere (i.e., it has no ``holes'' and thus can be deformed into a sphere without changing its topological structure) and consider a  closed curve $\ell $ on $\partial {\cal D}$ that splits $\partial {\cal D}$ into two separate parts. Then Eq.(\ref{alf3}) recovers  the classical result 
according to which  the flux through  a 2D surface bounded by a closed curve  $\ell$ is independent of the specific surface that is chosen and shows that  this result  is Lorentz invariant and  that, obviously,  it is conserved by the domain advection.  Clearly,  when the domain 
 ${\cal D} $ is Lorentz  boosted  to a moving frame,  it does not remain purely spatial (i.e., at constant time). 
 This lack of simultaneity in the busted frame can be corrected by the procedure of ``time resetting ''  mentioned above in Sec.\ref{FS}.

 \section{Conclusions}\label{concl}
 
 A first aim of this article was the completion of the Lorentz invariant formulation of the topological invariants  initiated in \cite{Peg1,Peg2},  where the concepts of covariant magnetic connectiond and of 2D connection hypersurfaces were introduced, by proving that the time resetting procedure can be applied in order to obtain a Lorentz invariant formulation of the Alfv\`en theorem for the conservation of the magnetic flux through a surface comoving with an ideal plasma.
 
 The procedure adopted in this derivation moves from the definition of a Lorentz invariant ``potential magnetic field''    similar to the potential vorticity in the non-relativistic Ertel theorem. As an additional bonus  this definition  makes it possible to set  in Lorentz invariant form the so called  Cauchy conditions for the magnetic field  components in the mapping  between Eulerian and Lagrangian variables. 

We remark that a  Lorentz invariant definition of the magnetic flux conservation in an ideal plasma is important  both  from a theoretical and in particular   from an experimental point of view. This is the case  for example when observing a relativistically  expanding plasma,  because    the distinction between electric and magnetic fields is frame dependent and  because simultaneity  between events  at different spatial locations is not maintained in  reference frames  moving with different velocities with respect to the observer.

Finally we notice that the results obtained in this article can be extended to non ideal plasmas that obey a generalised ideal Ohm's law, including e.g.,  electron inertia, as discussed explicitly  in  \cite{Peg3}. 
A possible extension to more general fluid theories (see e.g., \cite{Morr} in the non relativistic limit) and in particular to those that involve  an antisymmetric tensor  that unifies the electromagnetic and the fluid fields, \cite{asen1,asen2} should  also be investigated.
On the  contrary a possible extension to an ideal plasma in curved space-time must incude the fact that in General Relativity  covariant derivatives do not commute (which would result in a modification of Eq.(\ref{1})). 

\section*{Acknowledgment}

The author thanks Prof. P. J. Morrison for useful information.


\begin{thebibliography}{99}


\bibitem{Zan}  Del Zanna,  L.,  Pili,  A.G., Olmi,  B.,  Bucciantini,  N. \&  Amato, E.,  2018, {\it Plasma Phys. Control. Fusion}, {\bf 60},  014027.


\bibitem{Zen}  
 Zenitani, S.,   Hesse,  M. \& Klimas, A., 2010,  {\it ApJ.  Lett.}, {\bf 716 L},  214.


 \bibitem{Peg1} Pegoraro, F., 2012,  {\it EPL}, {\bf 99},  35001.
 
 
\bibitem{Peg2}  Pegoraro, F., 2016, 
{\it J. Plasma Phys}, {\bf  82} , 555820201.



 \bibitem{hide1}  Hide, R.,   1983,   
 {\it Ann. Geophys.},  {\bf 1} , 59.

 \bibitem{hide2}   Hide, R., 1996,  
{\it Geophys. J. Int.}, {\bf  125}, F1.

 \bibitem{ertel}   Ertel, H., 1942,  
 {\it Mereord. Zeit.},  {\bf 59}, 271.
 


 
\bibitem{newcomb}  Newcomb, W.A., 1962, {\it Nucl. Fus.  Suppl},  Part {\bf 2},   451.


\bibitem{frish} Frisch  U. \&   Villone,  B., 2014 
{\it The Euro. Phys. Jour. H}, {\bf 39}, 325.


\bibitem{Peg3}  Pegoraro, F.,  2015, {\it  Phys. Plasmas}, {\bf 22}, 112106.

\bibitem{Li}  Lichnerowicz,  A., 1967, in {\it Relativistic Hydrodynamics and Magnetohydrodynamics},
(New York: Benjamin) (1967).

\bibitem{Ani}  Anile, M., 1989, in    {\it Relativistic fluids and magneto-fluids}, Cambridge Monographs on
Mathematical Physics, Cambridge.

 \bibitem{D'A}  D'Avignon, E.,  . Morrison,  P.J.  \& Pegoraro, F.,  2015,  {\it Phys. Rev. D}, {\bf 91},  084050.


\bibitem{AMP-13}  Andreussi, T.,  Morrison  P.J., \& Pegoraro, F.,  2013, {\it Phys. Plasmas}, {\bf 20},  092104.
 
 
 \bibitem{Fitz}    Fitzpatrick  R., 2008, in {\it Maxwell's Equations and the Principles of Electromagnetism},  Jones \& Bartlett Learning Ed., $1^{st}$ edition Ch.10.24, (Boston MA).
 

 
 \bibitem{Morr}   
 Kawazura, Y.,  . Miloshevich,  G., \& Morrison  P.J., 2017, 
{\it Phys. f Plasmas}  {\bf 24}, 022103.

 \bibitem{asen1} Asenjo, F.A., \&  Comisso, L., 2015 , {\it Phys. Rev. Lett.}, {\bf  114}, 115003.


 \bibitem{asen2}    Asenjo,  F.A.,  Comisso, L. \& Mahajan, S. M. ,  2015, {\it Phys. Plasmas}, {\bf  22}, 122109


 
 
 \end{thebibliography}
\end{document}